\documentstyle[epsf]{mn}

\begin{document}

\title[Long-timescale X-ray variability of NGC~4051]
{The swansong in context: long-timescale X-ray variability of NGC~4051}
\author[P. Uttley et al.]
{P. Uttley$^{1}$\thanks{e-mail: pu@astro.soton.ac.uk}, 
I. M. M$^{\rm c}$Hardy$^{1}$, I. E. Papadakis$^{2}$, M. Guainazzi$^{3}$,
A. Fruscione$^{4}$  \\
$^{1}$Department of Physics and Astronomy, University of Southampton, 
Southampton SO17 1BJ \\
$^{2}$Physics Department, University of Crete, PO Box 2208, 710 03
Heraklion, Crete, Greece \\
$^{3}$Astrophysics Division, Space Science Department of ESA, ESTEC,
Postbus 299, 2200 AG Noordwijk, The Netherlands \\
$^{4}$Harvard-Smithsonian Center for Astrophysics, 60 Garden Street,
Cambridge, MA 02138, USA 
}

\date{Accepted; Received 1999 March 23}

\maketitle
\parindent 18pt
\begin{abstract} 
On 9-11 May 1998, the highly-variable, low luminosity Seyfert~1 galaxy
NGC~4051 was
observed in an unusual low flux state by {\it BeppoSAX} (Guainazzi et al. 1998)
{\it RXTE} and {\it EUVE}.  We present fits of
the 4--15~keV {\it RXTE} spectrum and {\it BeppoSAX} MECS
spectrum obtained
during this observation, which are consistent with the interpretation
that the source had switched off, leaving only the spectrum of pure
reflection from distant
cold matter.  We place this result in context by showing the X-ray
lightcurve of NGC~4051 obtained by our
{\it RXTE} monitoring campaign over the past two and a half years, which
shows that the low state lasted for $\sim150$~days before the May
observations (implying that the reflecting material is $>10^{17}$~cm from
the continuum source) and forms part
of a lightcurve showing distinct variations in long-term average flux over
timescales $>$~months.  We show that the long-timescale component to X-ray
variability is intrinsic to the primary continuum and is probably distinct
from the variability at shorter timescales, possibly associated with variations
in the accretion flow of matter onto the central black hole.  As the
source approaches the low state,
the variability process becomes non-linear.  NGC~4051 may represent a
microcosm of all X-ray variability in radio quiet active galactic nuclei
(AGNs), displaying in a few years a variety of flux states and variability
properties which more luminous AGNs may pass through on timescales of decades to
thousands of years.
\end{abstract}


\section{Introduction}

On 9-11 May 1998, the highly variable, low-luminosity (2--10~keV luminosity, 
$L_{2-10}\sim5\times 10^{41}$~ergs~s$^{-1}$)
Seyfert 1 galaxy NGC~4051 was observed in an extremely low, constant flux state
(2--10~keV flux $\sim1.3\times10^{-12}$~ergs~cm$^{-2}$~s$^{-1}$,
corresponding to $L_{2-10}\sim3\times 10^{40}$~ergs~s$^{-1}$) by
the Rossi X-ray Timing Explorer ({\it RXTE}), the Italian-Dutch X-ray
astronomy satellite {\it BeppoSAX} and the Extreme Ultraviolet Explorer
({\it EUVE}).  The {\it BeppoSAX} data were consistent with the
intepretation that the source had `switched off', leaving only the X-rays
reflected from distant cold matter (possibly the molecular torus) as a
witness to its earlier intensity (Guainazzi et al., 1998, henceforth
G98).

In this paper, we present the {\it RXTE} spectrum of the source in its
low state, confirming the interpretation presented in G98.  We
also place this `swansong' of the AGN in NGC~4051 in context, by showing
the two and a half year lightcurve of NGC~4051 obtained with {\it RXTE},
which shows a decline of the source average flux for nearly two
years, culminating in the low state which lasted $\sim150$~days
before the source `switched on' once more.

Variations on timescales of years, or even on $\sim150$~days, in NGC4051
are particularly interesting, given the general pattern of AGN X-ray
variability.  On short timescales (minutes--hours), AGN such as
NGC~4051 display scale invariant variability (e.g. M$^{\rm c}$Hardy \& Czerny
1987; Lawrence et al. 1987; M$^{\rm c}$Hardy 1988; Green,
M$^{\rm c}$Hardy \& Lehto 1993; Lawrence \& Papadakis 1993) which
can be seen from the power-law shape
of their X-ray power spectra.  However on longer timescales, $>$~day in the
case of NGC4051 (McHardy, Papadakis \& Uttley 1998) and $>$~month in the
case of the higher luminosity AGNs NGC~5506 (M$^{\rm c}$Hardy 1988)
and NGC~3516
(Edelson and Nandra 1999), the power spectra flatten, as they must if
the total variable power is not to become infinite.  Thus the 150~days
to few years timescale which we detect here is much longer than the
flattening or `knee' timescale in NGC~4051.  In section 5 we discuss
this result in the context of the mechanism for the long-timescale X-ray
variability and speculate on the implications for other AGN.

\section{Observations and Data Reduction}
For the past two and a half years we have monitored NGC~4051 with {\it RXTE} 
in order to investigate its variability across a broad range of timescales. 
To this end, we have used short ($<1$~ksec) observations to obtain
`snapshots' of the source flux through a range of time intervals.  From
May 1996 we observed the source twice daily for two weeks,
daily for four more weeks and at weekly intervals for the remainder of
the year.  Since 1997 we have observed the source every two weeks.  We
also observed NGC~4051 for a continuous period from
1998~May~9~16:43:12~UTC to 1998~May~11~20:55:28~UTC (61~ksec useful
exposure), simultaneous with observations by {\it BeppoSAX} and {\it EUVE}.

{\it RXTE} observed NGC~4051 with the Proportional Counter Array (PCA) and
the High
Energy X-ray Timing Experiment (HEXTE) instruments.  The PCA consists of 5
Xenon-filled
Proportional Counter Units (PCUs), sensitive to X-ray energies from
2--60~keV.  The HEXTE covers a range of between 20--200~keV, but due to
the faint nature of the source we only consider the PCA data in this work. 
Discharge problems mean that of the 5 PCUs
in the PCA, PCUs 3 and 4 are often switched off, so we include data from
PCUs 0, 1 and 2 only.  We extract data from the top layer of
the PCA using the standard {\sc
ftools 4.1} package, using the standard GTI criteria for
electron contamination and excluding data obtained within and up to 30 minutes
after
SAA maximum and data obtained with earth elevation $<10^{\circ}$.  We
estimate the background for the PCA with {\sc pcabackest v2.0c} using
the new L7 model for faint sources.
  
\section{The low state X-ray spectrum}

We now investigate the spectrum of the source in its low state, as 
measured by the PCA on board {\it RXTE} and the MECS instrument on {\it
BeppoSAX} during the long-look of May 9-11 1998. 
The 2-10~keV lightcurve obtained by the PCA shows no significant
variability above the expected level for systematic errors in the
background estimation, consistent with the observed lack of variability
in the {\it BeppoSAX} lightcurves (G98).  We therefore use the PCA and
MECS spectra integrated over the whole observation.

We will not consider the data from {\it EUVE} and the
LECS instrument on board {\it BeppoSAX} in our fits.  These data
show evidence for a separate low-energy component at energies below
4~keV, in addition to the component seen at medium energies by the
PCA and MECS.  This low-energy component is also constant in flux over
the 7 day duration of the {\it EUVE} observation (Fruscione, in preparation). 
Since we are interested in the medium energy spectral component, we
shall only consider
the PCA and MECS spectra in the energy ranges 4--15~keV and
4--10.5~keV respectively. 
We use a PCA response matrix generated by the {\sc pcarsp v2.36} script;
details of the MECS calibration and data reduction can be found in G98.

We fit the spectra in {\sc xspec v10.0}.  Simple power-law fits show a
very flat spectrum so, as in G98, we shall attempt to account for this
hard spectrum in terms of a reflection model.  By fitting up to 15~keV we can
use the simple {\bf href} multiplicative model for reflection of a power-law
spectrum off a slab of cold material.  We also include a gaussian
iron line and galactic absorption.  This simple model approximates
the reflection spectrum of cold material with an unknown distribution
around the primary X-ray source.  Like G98, we assume that the
reflecting material subtends $2\pi$ steradians of sky, as seen from the
source of the incident continuum.  The inclination angle of the reflector
to the line of sight
is unknown, but since it does not significantly affect the
fits, we freeze it arbitrarily at 30 degrees.  We find that the
best-fitting observed
fraction of the illuminating power-law continuum in all our model fits
where it is left free is zero (i.e. the source has switched off
completely); so we fix this parameter to zero for the purpose of
constraining the other model parameters. 
 
In table 1, we show the resulting best-fitting parameters for separate
model fits to the PCA and MECS spectra.  Both sets of data are fitted
reasonably well by the model and the model parameters are consistent with
being the same in both the PCA and MECS spectra.  The agreement between
the two instruments confirms the accuracy
of the PCA background model.  We therefore attempt
to constrain the model parameters further by fitting both the PCA and
MECS spectra jointly with the same model parameters.  The resulting
best-fitting parameters are also shown in table 1. 

\begin{table*}
 \caption{Best-fitting reflection model parameters for PCA, MECS and
          combined MECS and PCA spectra.}
 \label{symbols}
 \begin{tabular}{@{}lccccccc}
     & $\Gamma$ & $A$ & $E_{\rm K}$ & $\sigma_{\rm K}$ & $F_{\rm K}$ &
      $\chi^{2}/{\rm d.o.f.}$ & $L_{2-10}$/$10^{42}$~ergs~s$^{-1}$ \\
 PCA & $2.1\pm^{0.4}_{0.6}$ & $(0.6\pm^{1.0}_{0.5})\times10^{-2}$ &
      $6.42\pm^{0.18}_{0.17}$ & $0.29\pm^{0.47}_{0.29}$ &
      $(2.7\pm^{2.0}_{1.1})\times10^{-5}$ & $16.4/26$ & 0.30 \\
\\
 MECS & $2.2\pm^{0.4}_{0.5}$ & $(0.9\pm^{1.1}_{0.5})\times10^{-2}$ &
      $6.46\pm^{0.32}_{0.11}$ & $0\pm^{0.56}_{0}$ &
      $(1.60\pm^{1.2}_{0.6})\times10^{-5}$ & $28.2/22$ & 0.39 \\
\\
 PCA \& MECS & $2.30\pm0.25$ & 
      $(1.0\pm^{0.7}_{0.3})\times10^{-2}$ &
      $6.46\pm^{0.16}_{0.09}$ & $0.1\pm^{0.35}_{0.1}$ &      
      $(2.0\pm^{0.8}_{0.6})\times10^{-5}$ & $49.0/53$ & 0.38 \\
 \end{tabular}

\medskip
$\Gamma$ is the photon index of the incident power-law continuum, $A$ is
the incident power-law normalisation, $E_{\rm K}$ and $\sigma_{\rm K}$
are the line energy and actual width respectively (in keV) and $F_{\rm
K}$ is the line flux in photons~cm$^{-2}$~s$^{-1}$.  All errors are 90\%
confidence limits for 2 interesting parameters.  Also shown is the
$\chi^{2}$ value and number of degrees of freedom for each fit, and the
2--10~keV luminosity of the incident power-law (assuming $H_{0}=50$~km~s$^{-1}$~Mpc$^{-1}$).
\end{table*}
  
\begin{figure}
\begin{center}
{\epsfxsize 0.9\hsize
 \leavevmode  
 \epsffile{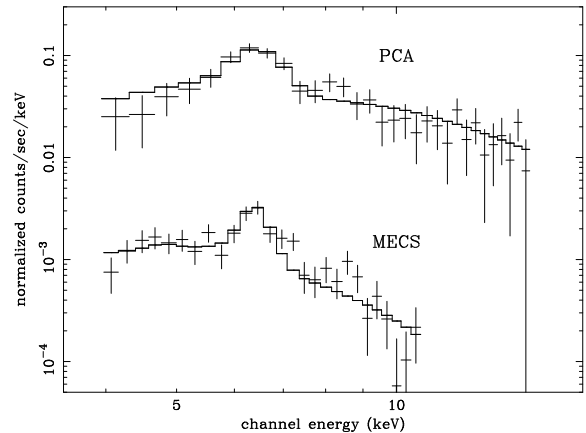}
}\caption{May 9-11 {\it RXTE} PCA and {\it BeppoSAX} MECS spectra for
the
best-fitting multiplicative reflection model described in the text.}
\end{center}
\label{fig:spec}
\end{figure}

Fig. 1 shows the model fitted jointly to the spectra from both
instruments. 
The inferred slope of the illuminating continuum, as obtained by the
joint fit, is higher than that
obtained from the individual fits to both the PCA and MECS spectra. 
The higher slope is due to the improved definition of
the continuum flux at lower energies by the MECS data, which sets the
1~keV continuum normalisation to a higher value than that given by the PCA fit
alone, combined with the greater sensitivity of the PCA at high energies
which holds down the continuum at higher energies.  As one might expect,
given that none of the illuminating continuum is directly visible, the
slope of that continuum is not well determined. 
However, we note that a continuum photon index of 2.3 was observed
during simultaneous {\it RXTE} and Extreme Ultraviolet Explorer ({\it
EUVE}) observations in May 1996 (Uttley et al., submitted to MNRAS).

The value inferred for the luminosity of the primary continuum incident
on the reflector is fairly
typical for NGC~4051 in its active state (e.g. Guainazzi et al., 1996).  Note
that the inferred value for the incident
luminosity assumes the slab geometry which is inherent in the
reflection model (i.e. 50\% covering fraction).  Since the inferred
continuum luminosity is compatible with observations, the actual
covering fraction must be of this order. 
Fixing the continuum slope of the combined-fit model to its
best-fitting value,
we can set a 99\% confidence upper limit (for 2 interesting parameters)
of 0.024 for the fraction of the illuminating primary flux which is
directly observed.  Combining this
observation with the lack of variability from the EUV to medium X-ray
bands, the simplest assumption is that the primary continuum has
switched off completely.

The iron line parameters are not strongly affected by the parametrization
of the underlying continuum.  The iron line equivalent width is
$\sim1$~keV, consistent with the interpretation that the entire
medium-energy spectrum originates in cold
reflecting material.  The line energy and width are also consistent with
this interpretation.
 
We conclude that the May 1998 long looks at NGC~4051 show that
the primary continuum source had switched off for at least seven days
leaving the clear signature of
reflected emission from cold material, possibly the molecular torus.  We
will now place this result in context by looking at the source history
for the two years preceding these observations and the six months
following them.

\section{long timescale variability}

In the upper panel of fig. 2 we show the two and a half year 2--10~keV
lightcurve of NGC~4051
obtained with our monitoring observations (crosses).  The May long look
observation is indicated by a star.  The lightcurve shows variability on
a range of timescales, including a probable long-term component on
timescales of $\sim$months, which
we  highlight in the lower panel of fig. 2 which shows the
100 day average fluxes, made with the monitoring data in the
corresponding 100 day bins.  The error bars on the 100 day averages are
intended to represent the spread of points in each bin and do not
represent an actual error on the 100 day mean.  The lightcurve
shows a decline from a high flux state in 1996, through an
intermediate flux state in 1997, culminating in the low state in
early 1998 which lasted for $\sim150$~days.  Shortly after the
long-look observations in May 1998 (and as far as the most recent
observations) the source became active again.

It is apparent from the lightcurve that the source flux variability is not
statistically stationary (i.e. has a constant mean) on long timescales.  We
now show that this long-term
variability is real and not some artifact due to the sparse sampling of
an underlying stationary, stochastic lightcurve.  We shall use only the
monitoring observations since they are all of comparable length and the
minimum seperation between observations is 0.5 days, which is of the
order of the knee timescale.  Hence if there are no
long-timescale components to variability we would expect any
lightcurve made up of such observations to be statistically stationary. 
\begin{figure*}
\begin{center}
{\epsfxsize 0.8\hsize
 \leavevmode  
 \epsffile{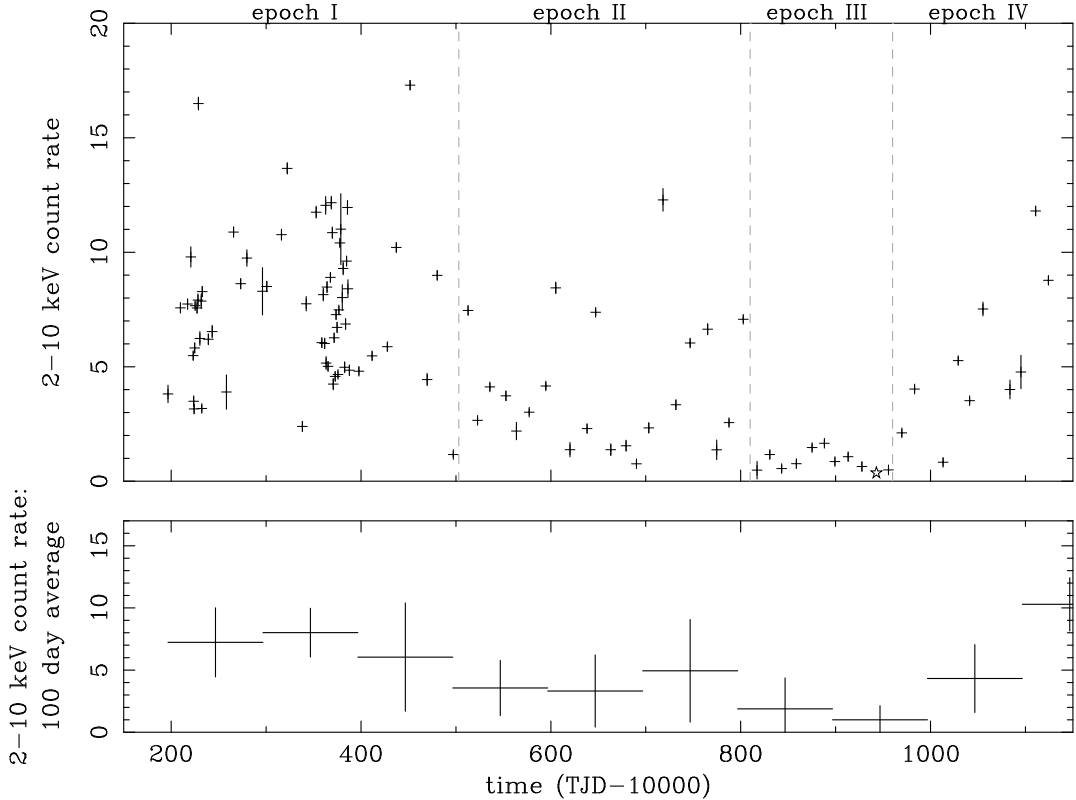}
}\caption{Two and a half year lightcurve of NGC~4051.  Errors in the top
panel are
$1\sigma$ counting errors.  Errors in the bottom panel are indicative
of the spread of points in each 100 day bin.  The May 1998 long-look
is represented by a
star.  Time units are Truncated Julian Date (TJD), minus $10^{4}$ for
presentation purposes.  Grey dashed lines demarcate the epochs described
in the text.}
\end{center}
\label{fig:lcurve}
\end{figure*}

We can compare the mean fluxes of two sections of the lightcurve using
Student's t-test for significantly different means
(e.g. Press et al., 1992), which gives a probability
that both sets of data have the same mean.  We first compare the mean
2--10 keV count rate of the 10 observations between TJD 10810 and 10860,
(when the source
appears to occupy an ultra-low state), with the mean count rate of the 
preceding 87 observations (0.92 counts~s$^{-1}$ versus 6.75
counts~s$^{-1}$).  The t-test shows that probability that both sets of
observations come from the same parent population is $10^{-26}$. 
However, we have selected the
group of low-state observations because they look different to the
preceding observations.  Therefore we must confirm the likelihood that
no group of 10 or
more consecutive observations can be found to be from the same
parent population as
a preceding group of observations (with a probability of less than or
equal to $10^{-26}$),
in any randomly generated
statistically stationary lightcurve with the same number of data points
as our own (106 observations).  We have simulated $10^{4}$
statistically stationary lightcurves each with 106 data points.  We then
searched each simulated lightcurve for groups of 10 or more consecutive
points with a
mean count rate which is the same as the mean of the preceding points with a
probability of $10^{-26}$ or less.  In $10^{4}$ lightcurves we find no
such groups, so we conclude that the mean of the lightcurve between TJD
10810 and 10960 is different to the mean in the preceding time period
at a level of better than $99.99\%$ confidence.  Therefore, the entire
lightcurve is not statistically stationary.

We now determine whether the lightcurve is statistically stationary prior to the
ultra-low state, i.e. does the source simply switch between two flux
states with constant average fluxes, or is there a more gradual change
in the mean flux, culminating in the low state?  To examine this
possibility, we split the lightcurve into two parts of equal duration,
corresponding to TJD 10196--10503 and TJD 10503--10810, with mean count
rates 7.64 and 4.19 respectively.  According to the t-test, these two
sections of the lightcurve have significantly different means at
better than $99.99\%$ confidence.  We cannot show that the lightcurves
are not stationary on shorter timescales by further splitting these
sections of the lightcurves into equal halves.  Therefore the X-ray lightcurve
of NGC~4051 approximates a stationary lightcurve on timescales of
days--weeks (hence the knee in the power spectrum), becoming non-stationary on
longer timescales.  We delineate these significantly different flux epochs
in Figure 2, naming them epochs I, II, III, and IV.  Epoch
IV is the most recent section of the lightcurve, where the source seems
to have returned to an active state.

Long-term variations in the average X-ray flux might be caused by
absorption by a varying column of material along the line of sight.  A
50\% reduction in average flux (i.e. between epoch I and epoch II)
requires an additional absorbing column of column density
$\sim10^{23}$~cm$^{-2}$, which is ruled out at greater than 99.9\%
confidence by our spectral data from epoch II observations. 
Therefore the long-term variations must be intrinsic to the primary
X-ray continuum.

Finally, we comment on evidence for non-linearity in the lightcurve. 
Green, M$^{\rm c}$Hardy \& Done (1999) used the method of searching for
asymmetry in the distribution of measured flux about the mean to show
that the variability of NGC~4051 was non-linear during a {\it ROSAT}
observation in November 1991, while an observation in November 1992
showed no evidence for non-linearity.  Using the same technique, we find
that the variability in epoch II is non-linear to 90\% confidence
although there is no evidence for non-linearity during epoch
I (there is not sufficient data to comment on the linearity during the
other epochs).  It is interesting to note that extrapolating the differing
mean X-ray fluxes in epochs I and II into the {\it ROSAT} band yield
{\it ROSAT} count rates
similar to those in November 1992 and November 1991 respectively
(assuming a simple power law model with photon index 2.3 and galactic 
absorption), implying that the non-linear behaviour of NGC~4051 may be
associated with the intermediate-flux state which  characterises epoch II.

\section{Discussion}
We now discuss the implications of these results for the interpretation
of the low state spectrum and the origin of the long-timescale
variability. \\
\indent Although the source appears to be quiescent during the
three days of the May 1998 long-look, there does appear to be some low
level of variability during epoch III, although the flux level remains
very low, implying that the source may not be entirely switched off for
the entire duration of epoch III. 
The low flux state lasts longer than $\sim150$~days,
but the reflection spectrum seen in the May 1998 long look, which occurs
at the end of the low state, is
consistent with reflection of a continuum with much higher flux.  We
infer that the reflecting matter lies at distances equal to or
greater than $\sim150$~light-days from the continuum source ($>$~few times
$10^{17}$~cm), confirming the interpretation of G98 that we have
directly detected the X-ray reflection spectrum from the molecular torus
in NGC~4051.  It is interesting to note that there is no detectable
signature of neutral hydrogen gas along the line of sight to the continuum source
in NGC~4051, over and above the expected galactic absorption \cite{mch},
whereas the detection of a reflection spectrum, almost certainly from
the surrounding torus, implies substantial
columns (greater than $10^{24}$~cm$^{-2}$) out of the line of sight. 
This result is in agreement with the standard AGN unification
scenario, where we expect substantial differences in the column density
along different lines of sight to the central source. \\
\indent Assuming a lower limit to the timescale for the observed long-term
variability of $\sim150$~days (i.e., comparable to the duration of the
low state) we see that the long-term variability timescale is
much longer than the knee timescale in NGC~4051 (by a factor $>$~50). 
We therefore speculate
that the long timescale component to variability may have an altogether
different origin to the variability at much shorter timescales.  The short
knee timescale of NGC~4051 implies a low black hole mass of order
$10^{5}$~M$_{\odot}$ if the knee timescale scales linearly with black
hole mass \cite{mch98}, consistent with the relatively low luminosity of
this AGN.  The observed long-term variability timescale is much
longer than the dynamical timescale for a black hole of this mass,
or the thermal or sound-crossing timescales associated with the accretion disk
which may fuel the AGN \cite{ed}.  However, the long-term variability
timescale is comparable with the
viscous timescale of an accretion disk \cite{tr}.  We therefore
speculate that the long-term X-ray variability of NGC~4051 is
related to variations in the accretion flow in the X-ray emitting region
close to the massive black hole. \\
\indent Recently, evidence has emerged for a long timescale component (relative to
the knee timescale) to the X-ray variability in the galactic black hole
candidate, Cyg~X-1 \cite{rao} on timescales $>10^{3}$~s.  The
long-term variability timescales in Cyg~X-1 and NGC~4051 imply
a scaling with luminosity (and possibly black hole mass), similar to
the scaling with the knee timescale.  We therefore speculate that long-term
X-ray variability in more luminous AGNs will occur on even longer
timescales than in NGC~4051 - from decades to centuries for typical
Seyfert galaxies (of $L_{X}\sim10^{43}$~ergs~s$^{-1}$),
to thousands of years for quasars.  The X-ray variability of NGC~4051
may represent a microcosm of the X-ray variability of all AGN. \\
\indent Finally, we note that X-ray experiments with low spectral bandwidth may
misclassify sources like NGC~4051, which have recently switched off, as
being heavily absorbed AGN.

\section{Conclusions}
We have shown that the X-ray spectrum of NGC~4051 in its low
state observed by {\it RXTE} and {\it BeppoSAX} in May 1998 is
consistent with reflection of the primary continuum off distant
($>150$~light-days) cold gas, which may be the molecular torus envisaged
by the AGN unification model.

We have shown that the X-ray lightcurve of NGC~4051 is not statistically
stationary over long timescales, and that during the course of our monitoring
campaign, the source does not simply switch between
two flux states, but moves from a highly variable (probably linear) high
flux state, through an intermediate variable (possibly non-linear) flux
state, to the low state where variability was minimal. 
Since May 1998 the source has become active once more.

The long-timescale component to X-ray variability is intrinsic to the
primary continuum (and not varying obscuration), and may be associated
with variations in the accretion flow of the putative accretion disk,
assuming a relatively small black hole of 10$^{5}$~M$_{\odot}$,
consistent with the low luminosity of this AGN.

The X-ray variability of NGC~4051 may represent a microcosm of all AGN 
variability, showing in only a few years a range of states and
behaviours which more luminous AGN may pass through on timescales of
decades to thousands of years. 

\subsection*{Acknowledgments}
We wish to thank the {\it RXTE} and {\it BeppoSAX} schedulers for
efficiently co-ordinating and supporting these observations.  PU
acknowledges financial support from the Particle Physics and Astronomy
Research Council, who also provided grant support to IM$^{\rm c}$H.  MG
acknowledges an ESA Research fellowship.  AF was supported by {\it AXAF}
Science Center NASA contract NAS 8-39073.

\bsp

\end{document}